\documentclass[accepted]{uai2024} 
                        
\usepackage[american]{babel}

\usepackage{natbib} 
\usepackage{mathtools} 
\usepackage{booktabs} 
\usepackage{tikz} 

\usepackage{times}
\usepackage[latin9]{inputenc}
\usepackage{multirow}
\usepackage{amsmath}
\usepackage{amssymb}
\usepackage{array}
\usepackage{xcolor}
\usepackage{verbatim}
\usepackage{algorithm}
\usepackage{algpseudocode}
\usepackage{float}
\usepackage{latexsym}

\providecommand{\tabularnewline}{\\}
\floatstyle{ruled}
\newfloat{algorithm}{tbp}{loa}
\providecommand{\algorithmname}{Algorithm}
\floatname{algorithm}{\protect\algorithmname}

\newtheorem{theorem}{Theorem}
\newtheorem{assumption}{Assumption}
\newcommand{\indep}{\perp \!\!\! \perp}
\newcommand{\Step}[1]{\algrenewcommand{\alglinenumber}[1]{Step ##1: } #1}

\newcommand{\be}{\begin{equation}}
\newcommand{\en}{\end{equation}}
\newcommand{\bea}{\begin{eqnarray}}
\newcommand{\ena}{\end{eqnarray}}
\newcommand{\ba}{\begin{array}}

\newcommand{\ea}{\end{array}}

\newcommand{\logit}{ {\mathrm{logit}}}

\newcommand{\E}{ {\mathbb{E}} }

\title{Inference for Optimal Linear Treatment Regimes in Personalized Decision-making}

\author[1]{\href{mailto:<ycheng26@ncsu.edu>?Subject=Your UAI 2024 paper}{Yuwen Cheng}{}}
\author[1]{\href{mailto:<syang24@ncsu.edu>?Subject=Your UAI 2024 paper}{Shu Yang}{}}

\affil[1]{%
    Statistics Dept.\\
    North Carolina State University\\
    Raleigh, NC, USA
}

\begin{document}
\maketitle

\begin{abstract}
  Personalized decision-making, tailored to individual characteristics,
is gaining significant attention. The optimal treatment regime aims
to provide the best-expected outcome in the entire population, known
as the value function. One approach to determine this optimal regime
is by maximizing the Augmented Inverse Probability Weighting (AIPW)
estimator of the value function. 
However, the derived treatment regime can be intricate and nonlinear, limiting their use. For clarity and interoperability, we emphasize linear regimes and determine the optimal linear regime by optimizing the AIPW estimator within set constraints. 

While the AIPW estimator offers a viable path to estimating the optimal regime, current methodologies predominantly focus on its asymptotic distribution, leaving a gap in studying the linear regime itself. However, there are many benefits to understanding the regime, as pinpointing significant covariates can enhance treatment effects and provide future clinical guidance. In this paper, we explore the asymptotic distribution of the estimated linear regime. Our results show that the parameter associated with the linear regime follows a cube-root convergence to a non-normal limiting distribution characterized by the maximizer of a centered Gaussian process with a quadratic drift. When making inferences for the estimated linear regimes with cube-root convergence in practical scenarios, the standard nonparametric bootstrap is invalid. As a solution, we facilitate the \citet{cattaneo2020bootstrap} bootstrap technique to provide a consistent distributional approximation for the estimated linear regimes, validated further through simulations and real-world data applications from the eICU Collaborative Research Database.

\end{abstract}

\section{Introduction}\label{sec:intro}

The application of personalized decision-making, which customizes decisions based on individual characteristics, is garnering significant interest across various fields such as economics \citep{behncke2009targeting, turvey2017optimal}, personalized medicine \citep{young2011comparative,zhang2020individualized}, and reinforcement learning \citep{jiang2016doubly, munos2016safe,fujimoto2019off}. 
Personalized medicine, in particular, tailors treatment decisions to the unique attributes of individual patients. A treatment regime takes a patient's specific characteristics as input and determines the appropriate treatment options as output. The optimal treatment regime (also known as an optimal policy, optimal strategy, individualized treatment
rule, etc) is the one that maximizes the overall benefit to the patient population, known as the value function.
This approach is closely related to a broad body of research in reinforcement learning. Evaluating the expected outcome of patients under a given treatment regime can be viewed as \textit{off-policy evaluation} (OPE) while identifying the optimal treatment regime that yields the highest value function can be considered \textit{off-policy learning} (OFL).


Numerous methods have been developed to identify optimal treatment
regimes. One category involves the regression-based approach, which
estimates the outcome mean function, referred to as the Q function.
It then determines the optimal regime based on this estimated Q function.
Techniques in this category include Q-learning \citep{watkins1992q,murphy2005generalization,qian2011performance}
and A-learning \citep{murphy2003optimal,shi2018high}, which estimate
the contrast function of outcome mean functions with varying treatments
instead of the original outcome mean function. Another further category
is policy search estimation, which is developed by maximizing the value function using either the Inverse Probability Weighting (IPW) estimator, the Augmented Inverse Probability Weighting (AIPW) estimator \citep{zhang2012robust,zhao2012estimating}, or the Targeted Minimum Loss-Based estimator (TMLE)
\citep{van2015targeted,luedtke2016super, montoya2021optimal, poulos2024targeted}.
In this paper, we focus on the policy search estimation for determining
the optimal regime, particularly highlighting the AIPW estimator.
This estimator is valued for its capability to incorporate machine
learning techniques for approximating nuisance parameters. Additionally,
it provides double robustness, meaning it remains consistent for the
value function if either the outcome mean function or the propensity
score model is accurately specified. 

Often, the resulting optimal
treatment regime can be complex and nonlinear, introducing high variability
and posing challenges for practical use. As a result,
it's a common practice to seek the optimal regime within a restricted class of regimes. A frequent subset is decision lists \citep{zhang2015using},  where regimes are sequences
of decision rules. Each rule is articulated as a series of if-then
statements that guide treatment recommendations. These regimes can be explored using tree-based methodologies \citep{doove2015novel,zhang2015using,zhao2015new}. In this paper, we focus on linear treatment regimes. These regimes not only pave the way for novel scientific discoveries and hypotheses but are also more accessible for both clinicians and patients.
The optimal
linear regime is determined by optimizing the AIPW estimator within the pre-specified linear regime constraints. 

While some recent papers have made important contributions to estimating
the treatment regimes, they focus on studying the asymptotic distributions
of the value function \citep{zhang2018interpretable,chu2023targeted}, 
and have not studied the asymptotic distribution of the estimated
treatment regimes. However, there are many benefits to understanding
the regime, as identifying significant covariates can enhance treatment
effects and offer insights for future clinical practices. To infer
the regime directly, it is valuable to determine the asymptotic distributions
of the estimated regime.  We prove that the estimated parameter indexing
the linear treatment regime converges at a cube-root rate to a non-normal
limiting distribution that is characterized by the maximizer of a
centered Gaussian process with a quadratic drift. Given its cube-root convergence, larger datasets are beneficial for real-world applications.
The rise of electronic health records (EHR) --- comprehensive digital patient histories maintained over
time --- offers vast datasets, encompassing demographics, clinical
notes, medical histories, and more. This means the requirement for
large samples can be readily met using resources like EHR. Hence our
study emphasizes large sample scenarios: in simulations, we use a
sample size of $20000$, and in real data analysis, we utilize the
eICU Collaborative Research Database (eICU-CRD) \citep{goldberger2000physiobank,pollard2018eicu,pollard2019eicu}
with $9697$ observations.

When constructing confidence intervals for the estimated linear regimes,
traditional bootstrap fails for the non-normal cube root convergence
estimators \citep{abrevaya2005bootstrap,leger2006bootstrap,cattaneo2020bootstrap}.
Several existing studies have presented consistent resampling-based
distributional approximations for cube-root-type estimators. One category
of methods achieves consistency by modifying the distribution used
to generate the bootstrap sample. This encompasses techniques like
subsampling \citep{seo2018local}, which uses without-replacement
subsamples to estimate the cube root estimator distribution; $m$
out of $n$ sampling \citep{lee2006m,bickel2012resampling} that employs
with-replacement samples for this estimation. Another category modifies
the objective function used  to construct the bootstrap-based distributional
approximation, and then applies the standard nonparametric bootstrap
to this altered function, as delineated by \citet{cattaneo2020bootstrap}.
In our research, we apply the bootstrap method proposed by \citet{cattaneo2020bootstrap}
to derive the confidence intervals, applying it to both simulation
and real data analysis, yielding promising results.

Our main contributions can be summarized as follows:
\begin{itemize}
    \item Building on the research of \citet{kim1990cube} and \citet{wang2018quantile} regarding mean-optimal and quantile-optimal criteria in randomized clinical trials, we establish the linear treatment regime for the AIPW estimator using observational data. We demonstrate that the estimated optimal treatment regime converges at a cube-root rate to a non-normal limiting distribution, characterized by the maximizer of a centered Gaussian process with quadratic drift.
   \item We utilize the bootstrap technique proposed by \citet{cattaneo2020bootstrap} to provide a consistent distributional approximation for the estimated optimal linear regime, addressing the challenges of applying standard nonparametric bootstrap methods to cube-root convergence estimators in practical scenarios.
   
\end{itemize}

The rest of the paper is organized as follows. In Section \ref{Background}, we present the basic
setup and introduce the AIPW estimator. Section \ref{sec:theorem} derives
the theorem of the asymptotic distributions of the AIPW estimator and the estimated linear treatment
regime. Additionally, this section introduces the algorithm based on the bootstrap method
introduced by \citet{cattaneo2020bootstrap} for the estimated linear
regimes. We conduct simulations in Section \ref{sec:Simulation}. Section
\ref{sec:Real-Data-Application} applies the proposed estimators to
an observational study from the eICU-CRD. Finally, we conclude the
paper with a discussion in Section \ref{sec:Discussion}.
\section{Background} \label{Background}

Denote $X\in\mathcal{X\text{\ensuremath{\subset\mathbb{R}^{l}}}}$
as the vector of pre-treatment covariates, $A\in\text{\{}0,1\text{\}}$
as the binary treatment, and $Y\in\mathbb{R}$ as the outcome of interest.
Following the potential outcomes framework, let $Y(a)$ be the potential
outcome for the subject given the treatment $a,a=0,1.$ The observed
data are then having $n$ independent and identically distributed
(i.i.d.) subjects $\left\{ \left(X_{i},A_{i},Y_{i}\right),i=1,\ldots,n\right\}.$

Consider a treatment regime, denoted as $d(X),$ that maps the $l$-dimensional
vector $X$ to the set $\{0,1\}.$ As an illustration, if we take $l = 1$ and define
the treatment strategy as $d(X)=I(X>0),$ then for $X=0.1,$ the assigned
treatment would be $1.$ Let $Y(d)$ represent the outcome an individual
would achieve when assigned a treatment based on regime $d(X):$
\[
Y(d)=Y(1)d(X)+Y(0)\left\{ 1-d(X)\right\}.
\]
 Further, denote the value function of regime $d(X)$ as \[V(d)=\E\left\{ Y(d)\right\} =\E\left[Y(1)d(X)+Y(0)\left\{ 1-d(X)\right\} \right],\]
which is the expected outcome under the regime $d(X).$ 
Assuming
our focus is on a specific collection $D$ of treatment regimes, the
optimal treatment regime is the one that maximizes the value function
$d^{opt}={\rm argmax}_{d\in D}V(d).$ For practical applications,
it's beneficial to consider that is both interpretable and straightforward
to implement, hence, in this paper, we focus on the linear regime
class $D_{\beta}=\{d(X;\beta)=I(X^{{\rm T}}\beta>0);\beta\in B\}$,
where different treatment regimes are indexed by $\beta$ and $B$
is a compact subset of the parameter space. Notably, $\beta$ is not
unique and is equivalent when scaled by scalar multipliers. Therefore,
we focus on $B=\{\beta,\|\beta\|=1\}$, where $\|\cdot\|$ denotes
the Euclidean norm. For brevity and clarity, within the context of
the linear regime class $D_{\beta},$ we represent its value function
$V(d)$ simply as $V(\beta).$ Denote $\beta_{0}={\rm {\rm argmax_{\beta\in B}}V(\beta),}$
then the optimal regime within regime $D_{\beta}$ is $d_{\beta}^{opt}\in D_{\beta}=d(X;\beta_{0})=I(X^{{\rm T}}\beta_{0}>0).$

One of the fundamental challenges to identifying the value function
is that $Y(1)$ and $Y(0)$ cannot be observed simultaneously. To
overcome this issue, we make the following three common assumptions
in the causal inference literature \citep{rubin1978bayesian}:

\begin{assumption}\label{asump-ignorable} $\{Y(0),Y(1)\}\indep A\mid X$
almost surely, where $\indep$ means ``independent of''. \end{assumption}

\begin{assumption}\label{consistency}$\ensuremath{Y=Y(1)A+Y(0)(1-A)}.$

\end{assumption}

\begin{assumption}\label{asump-overlap}There exist constants $c_{1}$
and $c_{2}$ such that $0<c_{1}\leq Pr(A\mid X)\leq c_{2}<1$ almost
surely.

\end{assumption}
Assumption \ref{asump-ignorable} tells us the assignment to treatment
is unconfounded. Assumption \ref{consistency}, known as the Stable Unit Treatment Value Assumption (SUTVA), suggests a lack of interference. This means that the potential outcomes for one individual remain unaffected by the treatments received or the potential outcomes of other individuals. Under Assumption \ref{asump-ignorable}-\ref{consistency}, the conditional mean of the potential outcome $Y(a)$ can be represented in terms of observed data. Specifically, the conditional outcome mean function $\mu_{A}\left(X\right)=\E\left\{ Y\left(A\right)\mid X\right\} =\E( Y\mid X,A)$.
Assumption \ref{asump-overlap} implies a sufficient overlap of the
covariate distribution between the treatment groups.

Denote the propensity score as $e(X)=Pr(A\mid X)$. Given the regime $d(X;\beta)$ and under Assumptions \ref{asump-ignorable}-\ref{asump-overlap}, it is imperative to highlight that \[\E \left[\frac{I\left\{ A=d(X;\beta)\right\} }{\rho(A\mid X)}\left\{ Y-\mu_{d}(X;\beta)\right\} \right]=0,\] where \begin{align*}
\rho(A\mid X) &=e(X)A+\left\{ 1-e(X)\right\} (1-A),\\
\mu_{d}(X;\beta) & =\mu_{1}(X)I\left\{ d(X;\beta)=1\right\} \\&+\mu_{0}(X)I\left\{ d(X;\beta)=0\right\}.
\end{align*}
Consequently, the value function can be expressed as
\begin{align*}
&V(\beta)\\=&\E\left[Y\{d(X;\beta)\}\right]\\
=&\E\left\{\mu_d(X;\beta) \right\}+\E \left[\frac{I\left\{ A=d(X;\beta)\right\} }{\rho(A\mid X)}\left\{ Y-\mu_{d}(X;\beta)\right\} \right].
\end{align*} 
Further, if we denote 
\begin{align*}
&v(X,A,Y;\beta)\\
=&\frac{I\left\{ A=d(X;\beta)\right\} }{\rho(A\mid X)}\left\{ Y-\mu_{d}(X;\beta)\right\} +{\mu}_{d}(X;\beta),
\end{align*}
then $V(\beta)$ can be written as $V(\beta)=\E \left\{v(X,A,Y;\beta) \right\}$.

Building upon this foundational understanding, \citet{zhang2012robust} proposed an AIPW estimator $\hat{V}_{n}(\beta)$ for the value function $V(\beta)$
as
\begin{align*}
\hat{V}_{n}(\beta)=\frac{1}{n}\sum_{i=1}^{n} \hat v(X_i,A_i,Y_i;\beta),
\end{align*} where
\begin{align*}
&\hat v (X_{i},A_{i},Y_{i};\beta)\\=&\frac{I\left\{ A_{i}=d(X_{i};\beta)\right\} }{\hat{\rho}(A_{i}\mid X_{i})}\left\{ Y_{i}-\hat{\mu}_{d}(X_{i};\beta)\right\} +\hat{\mu}_{d}(X_{i};\beta),
\end{align*}
and $\hat{e}(X),\hat{\mu}_{A}(X)$ are the estimates for $e(X)$ and $\mu_{A}(X)$. 
$\hat{\rho}(A\mid X)$ and $\hat{\mu}_{d}(X;\beta)$ are derived by substituting the estimates $\hat{e}(X)$ and $\hat{\mu}_{A}(X)$ into $\rho(A\mid X)$ and $\mu_{d}(X;\beta)$, respectively. Further, denote $\hat\beta={\rm{argmax}}_{\beta \in B}\hat{V}_n(\beta)$.

When using parametric models to estimate $\mu_A(X)$ (for $A=0,1$) or $e(X)$, the AIPW estimator consistently approximates $V(\beta)$ if either the posited parametric model for ${\mu}_A(X)$ (for $A=0,1$) or for ${e}(X)$ is correctly specified. Despite this advantageous property, real-world situations frequently pose difficulties in correctly implementing parametric models.
Recently, machine learning methods have gained traction.
 Various semi-parametric or non-parametric machine learning algorithms can be utilized to consistently estimate the unknown functions $e(X)$ and $\mu_{A}(X)$
for $A=0,1.$ Consequently, in this paper we focus on the use of semi-parametric or non-parametric machine learning models for estimating $e(X)$ and $\mu_{A}(X)$
for $A=0,1.$ Notably, these findings can be extended to the scenario where $e(X)$ and $\mu_{A}(X)$ are estimated via parametric models, and a detailed discussion is in Section \ref{sec:Discussion}.

\section{Methods}\label{sec:theorem}

In this section, we delve into several key topics. Subsection \ref{subsec:theorem for aipw} focuses on the derivation of the asymptotic distributions for the AIPW estimator. Meanwhile, subsection \ref{subsec:cube} explores the asymptotic distributions of
the estimated linear treatment regimes. Additionally, subsection \ref{bootstrap} introduces the algorithm
based on the bootstrap method proposed by \citet{cattaneo2020bootstrap}
for the estimated linear regime.

\subsection{Theorem}\label{subsec:theorem}
\subsubsection{Asymptotic distribution of the AIPW estimator }\label{subsec:theorem for aipw}

We begin
by outlining the theorem related to the asymptotic properties of $\hat{V}_{n}(\hat{\beta}).$
First, we assume the following regularity conditions:

\begin{assumption} \label{s1}The support of $X$ and $Y$ are bounded. 

\end{assumption}

\begin{assumption} \label{s2}The function $\mu_{a}(x)$ is smooth,
and continuously differentiable and bounded for all $(x,a).$

\end{assumption}

\begin{assumption} \label{s3}The optimal treatment regime $\beta_{0}\in B$
satisfying $\|\beta_{0}\|=1,$ is unique. 

\end{assumption}

\begin{assumption}\label{s4} 
\begin{align*}
&\left[\E\left\{ \hat{e}(X)-e(X)\right\} ^{2}\right]^{1/2}\sum_{a=0}^{1}\left[\E\left\{ \hat{\mu}_{a}(X)-\mu_{a}(X)\right\} ^{2}\right]^{1/2}\\=&o_{p}(n^{-1/2}).
\end{align*}

\end{assumption}

Assumptions \ref{s1}-\ref{s2} are standard regularity conditions
used to establish the convergence results. Assumption \ref{s3}
is an identifiability condition for $\beta_{0}$ and ensures the true
targeted optimal regime $d(X;\beta_{0})$ is uniquely defined, similar
to \citet{wang2018quantile}. To meet the criteria of Assumption \ref{s4}, one approach entails ensuring both $\sum_{a=0}^{1}\left[\E\left\{ \hat{\mu}_{a}(X)-\mu_{a}(X)\right\} ^{2}\right]^{1/2}=o_{p}(n^{-1/4})$ and $\left[\E\left\{ \hat{e}(X)-e(X)\right\} ^{2}\right]^{1/2}=o_{p}(n^{-1/4})$. In this context, purely nonparametric estimators, such as kernel or nearest-neighbor methods, are generally not viable due to their slower convergence rate, specifically below $o_p(n^{-1/4})$. However, certain semi-parametric models, like generalized additive models, can attain rates of $n^{-2/5}$. For a comprehensive list of estimators that can reach $o_p(n^{-1/4})$ convergence rates, we refer to the book \citep{horowitz2009semiparametric} and the review article \citep{kennedy2016semiparametric}.

The result is as
follows: 

\begin{theorem} \label{theorem1} Under Assumptions \ref{asump-ignorable}-\ref{s4},
as $n\rightarrow\infty$, we have 
\begin{enumerate}
\item $\|\hat{\beta}-\beta_{0}\|=O_{p}(n^{-1/3}).$
\item $\text{\ensuremath{\sqrt{n}}}\left\{ \hat{V}_{n}(\hat{\beta})-V({\beta_{0})}\right\} \stackrel{D}{\rightarrow}\mathcal{N}(0,\sigma^{2})$,
\end{enumerate}
where $\stackrel{D}{\rightarrow}$ represents convergence in distribution, and 
\begin{equation*}
\begin{split}
\sigma^{2}&=\E\Biggl[\frac{I\left\{ A_{i}=d(X_{i};\beta_{0})\right\} }{\rho(A_{i}\mid X_{i})}\left\{ Y_{i}-\mu_{d}(X_{i};\beta_{0})\right\}\\ &+\mu_{d}(X_{i};\beta_{0})-V(\beta_{0})\Biggl]^{2}.
\end{split}
\end{equation*}
\end{theorem}

The asymptotic distribution results provide valuable insights for making inferences regarding $V(\beta_0)$. It's important to highlight that while $\hat{V}_{n}(\hat{\beta})$
converges at a $\sqrt{n}$-consistent rate, the convergence rate of
regime $\hat{\beta}$ is $n^{1/3}$. This deviates from many established
statistical theorems, such as the central limit theorem, which operates
on the square root rate $O_{p}(n^{-1/2})$. Despite the extensive
exploration of the $\sqrt{n}$-consistent AIPW estimator $\hat{V}_{n}(\hat{\beta})$
and its various modified versions, the asymptotic distribution of
$\hat{\beta}$ remains relatively unexplored due to its $n^{1/3}$
convergence rate. In the following section, we explore the asymptotic distribution of
$\hat{\beta}$.\textbf{ }\citet{wang2018quantile}\textbf{ }investigated
the linear treatment regime for both the mean-optimal and quantile-optimal
criteria in randomized clinical trials. We build upon their findings
to deduce the linear treatment regime for the AIPW estimator using
observational data.

\subsubsection{Asymptotic distribution of the estimated linear regime}\label{subsec:cube}
\citet{kim1990cube} deduced the asymptotic distribution related to
cube root convergence, however, the result of \citet{kim1990cube}
is not directly transferrable because $\hat{V}_{n}(\hat{\beta)}$
incorporates the estimated $\hat{e}(X)$ and $\hat{\mu}_{A}(X)$ for
$A=0,1.$ Hence, we examine the conditions outlined in \citet{kim1990cube}
sequentially in the proof of Theorem 2, as detailed in the Supplementary
material. First, we introduce the following conditions:

\begin{assumption} \label{s5}$X$ has a continuously differentiable
density $f(\cdot)$ and that the angular components of $X,$ considered
as a random element of the unit sphere $S$ in $R^{l}$, has a bounded
continuous density with respect to surface measure on $S.$ 

\end{assumption}

\begin{assumption}\label{s6}
\begin{align*}
    H&=\int\left\{ x^{{\rm T}}\beta_{0}=0\right\} \left\{ \dot{f}(x)h(x)+f(x)\dot{h}(x)\right\} ^{{\rm T}}\beta_{0}xx^{{\rm T}}d\sigma,
\end{align*}
and $H>0$, where $\sigma$ is the surface measure on the hyperplane $\{X:X^{{\rm T}}\beta_{0}=0\}$
and $h(x)=\mathbb{E}\left\{ Y(1)-Y(0)\mid X=x\right\}$. $\dot{f}(x)$ and $\dot{h}(x)$ denote the first-gradient with
 respect to $x$.

\end{assumption}

\begin{assumption}\label{s7}\textbf{$\sup_{x\in\mathcal{X}}\vert\hat{e}(X)-e(X)\vert=o_{p}(n^{-1/3}),$
}and\textbf{ $\sup_{x\in\mathcal{X}}\vert\hat{\mu}_{A}(X)-\mu_{A}(X)\vert=o_{p}(n^{-1/3})$
}for\textbf{ $A=0,1.$}

\end{assumption}
Assumptions \ref{s5}-\ref{s6} are technical conditions for evaluating
the first and second-order derivatives of the value function and the
kernel covariance which are used to characterize the asymptotic distribution
of $\hat{\beta},$ similar in Example 6.4 in \citet{kim1990cube}
and \citet{wang2018quantile}. Assumption \ref{s5} aids in deriving
the first derivative of $V(\beta_{0})$. Under Assumption \ref{s6},
as deduced from the proof of Theorem \ref{theorem2}, the matrix $-H$ represents
the second order derivative of value function $V(\beta_{0})$ at $\beta=\beta_{0}$,
\begin{align*}
H&=\int\left\{ x^{{\rm T}}\beta_{0}=0\right\} \left\{ \dot{f}(x)h(x)+f(x)\dot{h}(x)\right\} ^{{\rm T}}\beta_{0}xx^{{\rm T}}d\sigma\\&=-\partial^{2}V(\beta_{0})/\partial\beta\partial\beta^{{\rm T}}.
\end{align*}
Ensuring $H$ is positively definite is one crucial condition in \citet{kim1990cube}\textbf{
}for the asymptotic distribution of $n^{1/3}(\hat{\beta}-\beta_{0}).$

Assumption \ref{s7} requires $\hat{e}(X)$ and $\hat{\mu}_{A}(X)$
to uniformly converge to $e(X)$ and $\mu_{A}(X)$ at a rate of $o_{p}(n^{-1/3}).$ It's noticeable
that the requisite convergence rate of $\hat{e}(X)$ and $\hat{\mu}_{A}(X)$
for $A=0,1$ under Assumption \ref{s7} is faster than the approach suggested following Assumption \ref{s4}. Technically speaking, the reason for the faster $o_p(n^{-1/3})$ uniform convergence rate in Assumption \ref{s7} is to meet the first condition set out in \citet{kim1990cube}, ensuring the existence of the cube root asymptotic distribution of $\hat\beta$ exists.
A perspective to understand this is by recognizing
that when developing the $\sqrt{n}$-consistent rate asymptotic distribution of the
value function $\hat{V}_{n}(\hat{\beta}),$ it is only requiring the
$\|\hat{\beta}-\beta_{0}\|=O_{p}(n^{-1/3}),$ however, given $\|\hat{\beta}-\beta_{0}\|=O_{p}(n^{-1/3}),$
it still needs further restriction to guarantee the asymptotic distribution
of $\hat{\beta}.$ That's why the conditions are more strict in
Assumption \ref{s7} than the approach suggested following Assumption \ref{s4}. 

To guarantee that there are estimators reaching this $o_{p}(n^{-1/3})$
uniform convergence rate for estimating $\hat{e}(X)$ and $\hat{\mu}_{A}(X)$
for $A=0,1$, first assuming that both $e(X)$ and $\mu_{A}(X)$ for
$A=0,1$ belong to the function class $\Sigma_{s},$ where $\Sigma_{s}$
represents the Holder classes of smoothness order $s$. For $s\in (0,1]$, the Holder class $\Sigma_{s}$ is defined as the set of all functions $f$: $\mathcal{X}\rightarrow \mathbb{R}$ such that for $C > 0$,
\[
\vert f(x)-f(\tilde{x})\vert \leq C \left\{\sum_{j=1}^{l}(x_{j}-\tilde{x}_{j})^{2}\right\} ^{s/2}
\]
for all $x,\tilde{x}\in\mathcal{X}$. For $s>1$, $\Sigma_{s}$ is defined as follows. For any $\alpha=(\alpha_1,\cdots,\alpha_l)$ of nonnegative integers, denote $D^\alpha=\partial_{x_1}^{\alpha_1}\cdots\partial_{x_l}^{\alpha_l}$.
Then $\Sigma_{s}$
is the set of all functions  $f:$ $\mathcal{X}\rightarrow \mathbb{R}$ such that $f$ is $[s]$ times continuously differentiable
and for some $C>0$, \[\vert D^{\alpha}f(x)-D^{\alpha}f(\tilde{x})\vert\leq C\left\{ \sum_{j=1}^{l}(x_{j}-\tilde{x}_{j})^{2}\right\} ^{(s-[s])/2}\]
and $\vert D^{\beta}f(x)\vert\leq C$ hold for all $x,\tilde{x}\in\mathcal{X}$,
where $\alpha=(\alpha_{1},\ldots,\alpha_{l})$ and $\beta=(\beta_{1},\ldots,\beta_{l})$
are nonnegative integers satisfying $\alpha_{1}+\cdots\alpha_{l}=[s]$
and $\beta_{1}+\cdots\beta_{l}\leq[s]$. Given these assumptions,
the optimal uniform rate of convergence for $\hat{e}(X)$ and $\hat{\mu}_{A}(X)$
(for $A=0,1$) is $O_{p}\left\{ (\ln n/n)^{s/(2s+l)}\right\} $ \citep{stone1982optimal}.
This rate can be achieved using various estimators, such as series
estimators \citep{belloni2015some} and local polynomial (kernel)
estimators \citep{takezawa2005introduction}. 
Additionally, when $e(X)$ and $\mu_A(X)$ (for $A=0,1$) are determined using estimators that meet this optimal uniform rate and if $e(X)$ and $\mu_A(X)$ adhere to the condition $s>l$, the uniform rate achieves $o_{p}(n^{-1/3})$.
In such scenarios, it's feasible to employ estimators that realize this $o_p(n^{-1/3})$ rate to estimate $e(X)$ and $\mu_{A}(X)$, where $A=0,1$.

The theorem is as follows: 

\begin{theorem}\label{theorem2}Under Assumptions \ref{asump-ignorable}-\ref{s3},
and Assumptions \ref{s5}-\ref{s7}, we have
\[
n^{1/3}(\hat{\beta}-\beta_{0})\stackrel{D}{\rightarrow}{\rm argmax}_{t}\left\{-\frac{1}{2}t^{{\rm T}}Ht+W(t)\right\},
\]
 where $\stackrel{D}{\rightarrow}$ represents converge in distribution. $H=-\partial^{2}V(\beta_{0})/\partial\beta\partial\beta^{{\rm T}}$
is a $l\times l$ positively definite matrix and $W(t)$ is a zero-mean
Gaussian process with continuous sample paths and covariance kernel
$C(\cdot,\cdot).$ The expressions for $C(\cdot,\cdot)$
is in the Supplementary material. 

\end{theorem}
\subsection{Bootstrap Algorithm}\label{bootstrap}
The distribution result in Theorem \ref{theorem2} sheds light on constructing inference on $\beta_0$ as we develop in subsection \ref{subsec:cube}. 
While we theoretically determine the asymptotic distribution of $\hat\beta$, applying these theoretical results for inference in practice can be challenging. A straightforward approach in real-world scenarios might involve using bootstrap methods to sample the distribution of $\hat\beta$. However,
the standard nonparametric bootstrap is often inadequate in approximating
the cube root distribution. We provide a straightforward demonstration
below. Let's define $V_n(\beta)=1/n\sum_{i=1}^n v(X_i,A_i,Y_i;\beta)$. According to \citet{kim1990cube}, the cube root convergence for $\hat{\beta}$ can be written as
$n^{1/3}(\hat{\beta}-\beta_{0})={\rm argmax}_{t\in \mathbb{R}^l}\{\hat{W}(t)+\mathcal{V}(t)\},$
\begin{equation*}
\begin{split}
\hat{W}(t)=n^{2/3}\Biggl\{ &{V}_{n}(\beta_{0}+tn^{-1/3})-{V}_{n}(\beta_{0})\\-&V(\beta_{0}+tn^{-1/3})+V(\beta_{0})\Biggl\}
\end{split}
\end{equation*}
is a zero-mean random process and asymptotically converges to $W(t)$
and 
\[
\mathcal{V}(t)=n^{2/3}\left\{ V(\beta_{0}+tn^{-1/3})-V(\beta_{0})\right\} 
\]
 asymptotically converges to $-t^{{\rm T}}Ht/2,$ where $H=-\partial^{2}V(\beta_{0})/\partial\beta\partial\beta^{{\rm T}}.$
The standard nonparametric bootstrap can replicate the shape of $\hat{W}(t)$,
however, it fails to replicate the shape of $\mathcal{V}(t)$, which
results in the inconsistency of the standard nonparametric bootstrap
\citep{abrevaya2005bootstrap,leger2006bootstrap,cattaneo2020bootstrap}.

To address this, a refined bootstrap methodology was proposed by \citet{cattaneo2020bootstrap}
to more accurately approximate the distribution for $\hat{\beta}.$
This method alters the objective function to ensure that the bootstrap
version of each empirical process counterpart has a mean resembling
its large sample version. Specifically, \citet{cattaneo2020bootstrap} reshaped the original objective
function $\hat v(X_{i},A_{i},Y_{i};\beta)$
to 
\begin{align*}
&\tilde{v}(X_{i},A_{i},Y_{i};\beta)\\=&\hat v(X_{i},A_{i},Y_{i};\beta)-\hat{V}_{n}(\beta)-\frac{1}{2}(\hat{\beta}-\beta)^{{\rm T}}H_{n}(\hat{\beta}-\beta),
\end{align*}
 where $H_{n}$ serves as an approximation of $H.$ This adjustment
ensures the convergence of the bootstrap versions to their population
counterparts in large samples. For $\hat{\beta}$, the bootstrap samples are represented as $\hat{\beta}^{*}$, given by
\begin{align*}
  \hat{\beta}^{*} &={\rm{argmax}}_{\beta \in B} \hat{V}_{n}^{*}(\beta),\\\hat{V}_{n}^{*}(\beta) &=\frac{1}{n}\sum_{i=1}^{n}\tilde{v}(X_{i}^{*},A_{i}^{*},Y_{i}^{*};\beta),
\end{align*}
 where $\left\{ (X_{i}^{*},A_{i}^{*},Y_{i}^{*}),i=1,\ldots,n\right\}$ are random samples from the empirical distribution $(X,Y,A)$. It's important to highlight that during this bootstrap procedure, the nuisance parameters can either be refitted using the samples $\left\{(X_{i}^{*},A_{i}^{*},Y_{i}^{*}),i=1,\ldots,n\right\}$ during the bootstrap process or retain their initially estimated values from the original datasets. Both approaches meet the conditions outlined in \citet{cattaneo2020bootstrap}. 
\citet{cattaneo2020bootstrap} proved that under certain regularity conditions, $n^{1/3}(\hat{\beta}^{*}-\hat{\beta})\rightarrow{\rm argmax}_{t}\{-\frac{1}{2}t^{{\rm T}}Ht+W(t)\}$ in distribution, therefore guarantees consistency of the bootstrap samples. Algorithm
\ref{alg:The-bootstrap-algorithm} outlines the detailed procedure
for bootstrapping samples. 
Regarding computational requirements, the complexity of this process is expressed as $O(K\tilde{B})$, where $\tilde{B}$ is the size of bootstrap samples, and $K$ denotes the algorithm's complexity for obtaining the estimate $\hat{\beta}$ given $\hat{V}_n(\beta)$. Specifically, employing a genetic algorithm \citep{katoch2021review} introduces a complexity of $K=O(GNn)$, with $G$ indicating the number of iterations and $N$ the population size.
Incorporating this methodology, we utilized
it to determine the 95\% confidence interval for $\hat{\beta}$ in
our simulations. 

\begin{algorithm}

\begin{algorithmic}[1]
\Step \State Using the sample $(X_{i},Y_{i},A_{i})$, compute $\hat{\beta}$ by approximately maximizing $\hat{V}_{n}(\beta)$. 
\Step \State Using $\hat{\beta}$ and $(X_{i},Y_{i},A_{i})$, compute $H_{n}$, where each $(k,l)$ element in $H_{n}$ is defined as: 
\begin{align*} 
H_{n,kl} &= -\frac{1}{4\epsilon_{n}^{2}}\bigg\{ \hat{V}_{n}(\hat{\beta}+e_{k}\epsilon_{n}+e_{l}\epsilon_{n}) \\& -\hat{V}_{n}(\hat{\beta}+e_{k}\epsilon_{n}-e_{l}\epsilon_{n}) -\hat{V}_{n}(\hat{\beta}-e_{k}\epsilon_{n}+e_{l}\epsilon_{n})\\
&+\hat{V}_{n}(\hat{\beta}-e_{k}\epsilon_{n}-e_{l}\epsilon_{n})\bigg\}, 
\end{align*} where $e_{k}$ is the $k$-th unit vector in $\mathbb{R}^{l}$ and $\epsilon_{n}$ is a positive tuning parameter. $H_{n}$ is a consistent estimator of $H$. 
\Step \State Using $\hat{\beta}$, $H_{n}$, and the bootstrap sample $\left\{ (X_{i}^{*},A_{i}^{*},Y_{i}^{*}),i=1,\ldots,n\right\}$, compute $\hat{\beta}^{*}$ by approximating maximizing $\hat{V}_{n}^{*}(\beta)$. 
\Step \State Repeat Step 3 to generate draws from the distribution $n^{1/3}(\hat{\beta}^{*}-\hat{\beta})$. 
\end{algorithmic} 

\caption{\label{alg:The-bootstrap-algorithm}The proposed bootstrap algorithm.}
\end{algorithm}

A key element of their approach revolves around the tuning parameter, $\epsilon_{n}$. Although \citet{cattaneo2020bootstrap} suggested an optimal value for $\epsilon_{n}$ that minimizes the approximate Mean Squared Error, this ideal value incorporates both the $H$ matrix and the covariance kernel $C(\cdot,\cdot)$. Determining the best $\epsilon_{n}$ involves estimating both $H$ and $C(\cdot,\cdot)$, a process that is intricate and not feasible in our context.

Beyond this method, there exist other strategies that guarantee consistency for cube root convergence estimators by adjusting
the distribution from which the bootstrap sample is drawn. Examples include subsampling methods, \citep{seo2018local}, which use subsamples without replacement, and the $m$ out of $n$ sampling techniques  \citep{lee2006m,bickel2012resampling}, which rely on samples with replacement for their estimates. \citet{hong2020numerical} introduced a numerical bootstrap technique where bootstrap samples are determined by the maximizer of the linear combination of the empirical distribution and the bootstrapped empirical process.

However, both the $m$ out of $n$ approach and subsampling necessitate that the count of bootstrap samples be $o_p(n)$. In practical applications, when maximizing a non-regular objective function to obtain the cube root convergence estimator, a limited count in the bootstrap samples can lead to results that deviate from true values, compromising the outcomes. The numerical bootstrap method \citep{hong2020numerical}, on the other hand, is also contingent on a tuning parameter.

\section{Simulation}\label{sec:Simulation}

In this section, we conduct one simulation study with $T=100$ simulation
times. To effectively achieve the cube root convergence in finite
samples, a substantial number of observations is essential. Thus,
in each simulation time, we generate $n=20000$ observations $(X_{i},Y_{i},A_{i}),i=1,\ldots,n,$
where $X_{i}=(1,X_{i1},X_{i2})^{{\rm T}}$ and $X_{i1}$ and $X_{i2}$
are independent following the uniform distribution on $\left[1-\sqrt{3},1+\sqrt{3}\right]$;
given $X_{i}$, the binary treatment indicator $A_{i}$ satisfies
$\logit\{e(X_{i})\}=-1.0+0.8X_{i1}+0.8X_{i2}$, where $\logit(u)=\log\{u/(1-u)\}$;
and outcomes are generated as $Y_{i}=2-1.5X_{i1}-1.5X_{i2}+A\times(2X_{i1}+X_{i2})+\epsilon_{i},\ \epsilon_{i}\sim\mathcal{N}(0,1)$
. To estimate $\logit\{e(X)\}$ and $\mu_{A}(X)$ for $A=0,1$, we
apply the generalized additive model (GAM) using the cubic smoothing
spline for each univariate covariate. From \citet{eggermont2001maximum},
the\textbf{ }cubic smoothing spline can achieve the optimal uniform
rate, therefore $\hat{e}(X)$ and $\hat{\mu}_{A}(X)$ (for $A=0,1$)
can uniformly converge to $e(X)$ and $\mu_{A}(X)$ for $A=0,1$ with
the rate $o_{p}(n^{-1/3}).$ 

For each individual, the optimal treatment regime is given by $I\left\{ \mu_{1}(X_{i})>\mu_{0}(X_{i})\right\},$
therefore in our example the optimal regime is $d(X_{i})=I(2X_{i1}+X_{i2}>0),$
which is a linear regime. The optimization process is facilitated
using the $\texttt{genoud}$ function in the R package $\texttt{rgenoud}$
\citep{mebane2011genetic}. To achieve the uniqueness, we impose the
restriction $\|\beta\|=1$, and the true optimal linear rule $\beta_{0}=(\beta_{01},\beta_{02})^{{\rm T}}=(0.894,0.447)^{{\rm T}}.$

To determine the 95\% confidence interval for $\hat{\beta},$ we incorporate
the \citet{cattaneo2020bootstrap} Bootstrap method, with the use
of 400 bootstrap samples. In each bootstrap sample, we refit both $e(X)$ and $\mu_A(X)$ to obtain the AIPW estimator and derive the estimated linear regimes. 
Identifying the optimal tuning parameter $\epsilon_{n}$, which minimizes the approximate Mean Squared Error, requires the estimated $H$ matrix
and\textbf{ }the covariance kernel $C(\cdot,\cdot).$ Considering the intricate nature and challenges presented in this context, we evaluate
multiple $\epsilon_{n}$ values to identify the most suitable one.

Table \ref{tab:simulation} presents the estimates for $\text{\ensuremath{\beta_{01}}}$
and $\beta_{02}$ (denoted as ``Est'') and their 95\% quantile confidence
interval length (denoted as ``Length'') and coverage rate (denoted
as ``Coverage''). The findings suggest that an optimal $\epsilon_{n}$
is approximately 0.5, as it achieves a 95\% coverage rate and the
shortest confidence interval length. This indicates that, with a judicious
selection of $\epsilon_{n}$, the \citet{cattaneo2020bootstrap} bootstrap
method can present a reasonable inference of the $\hat{\beta}.$
\begin{table}
\center{}
\caption{\label{tab:simulation}Simulation results under different tuning parameters
$\epsilon_{n}$ based on 100 Monte Carlo times with 400 bootstrap samples
in each simulation time. }
\setlength{\tabcolsep}{0.5mm}{
\begin{tabular}{ccccccccc}
\hline 
 & \multirow{2}{*}{Est} &  & \multicolumn{6}{c}{$\epsilon_{n}$}\tabularnewline
 &  &  & 0.05 & 0.1 & 0.2 & 0.5 & 0.7 & 0.9\tabularnewline
\hline 
\multirow{2}{*}{$\beta_{01}$} & \multirow{2}{*}{0.895} & Coverage  & 0.750 & 0.900 & 1 & 0.95 & 1 & 1\tabularnewline
 &  & Length & 0.139 & 0.894 & 0.984 & 0.086 & 0.135 & 0.235\tabularnewline
\multirow{2}{*}{$\beta_{02}$} & \multirow{2}{*}{0.443} & Coverage  & 0.740 & 0.920 & 1 & 0.95 & 1 & 1\tabularnewline
 &  & Length & 0.207 & 0.864 & 1.170 & 0.176 & 0.261 & 0.416\tabularnewline
\hline 
\end{tabular}}

\end{table}

\section{Real Data Application}\label{sec:Real-Data-Application}

We demonstrate our proposed approach using data sourced from the eICU-CRD,
a multi-center repository of anonymized health records spanning across
the United States from 2014 to 2015 \citep{goldberger2000physiobank,pollard2018eicu,pollard2019eicu}.

We consider the $9$ baseline covariates: age (years), Body Mass Index
(BMI), derived by dividing admission weight (kg) by the square of
admission height (meters), admission temperature (Temp) value (Celsius),
glucose level (mg/dL), blood urea nitrogen (BUN) amount (mg/dL), creatinine
amount (mg/dL), white blood cell (WBC) count (K/uL), bilirubin (mg/dL),
mean blood pressure (BP) level (mmHg). A treatment value of 1 indicates
the patient was administered vasopressor, while a value of 0 suggests
other medical interventions. We consider the cumulative balance (mL)
as the outcome of interest. A positive cumulative balance means the
fluid intake exceeds the output, leading to a condition called hypervolemia
or fluid overload. Excess fluid can strain the heart, potentially
causing heart failure\textbf{ }\citep{gologorsky2020ultrafiltration},
rapid decline in kidney function, and an increased need for kidney
replacement therapy \citep{palmer2020fluid}. Conversely, a negative
balance implies the patient's output exceeded their intake, labeled
as hypovolemia or fluid deficit. Severe hypovolemic shock can result
in mesenteric and coronary ischemia that can cause abdominal or chest
pain \citep{taghavi2018hypovolemic}. For our study, we use $Y=-\vert$cumulative
balance$\vert$ (CB) as the outcome, where a higher value is preferable.


After filtering the abnormal values, a total
of $9697$ observations remained. Table \ref{tab:summary data} summarizes
the mean and the standard deviation of the outcome and covariates
in the samples. 
To utilize the bootstrap method proposed by \citet{cattaneo2020bootstrap}, we generate 100 bootstrap samples to derive the 95\% confidence intervals for the estimated linear regimes and conduct sensitivity analysis across various $\epsilon_{n}$ values, specifically within the set $\{0.3,0.5,0.7\}$. 
Due to the widest confidence interval ranges observed at $\epsilon_n=0.7$, which results in no findings for significant covariates, we focus on presenting results for $\epsilon_n=0.3$ and $0.5$ in the main text. The corresponding results for $\epsilon_n=0.7$ are included in the supplementary materials. Detailed estimates and confidence intervals for $\epsilon_n=0.3$ and $0.5$ are shown in Table \ref{tab:realdata1}, where the intercept is denoted as ``Int''."
\begin{table}
\center{}%
\caption{\label{tab:summary data}Mean and the standard deviation (denoted
as ``sd'') of the outcome and covariates in the samples. }
\setlength{\tabcolsep}{0.5mm}{
\begin{tabular}{cccccc}
\hline 
 & age & BMI & Temp & Glucose & BUN\tabularnewline
\hline 
mean & 64.95 & 28.94 & 36.04 & 149.73 & 29.55 \tabularnewline
sd & 15.32 & 7.63 & 4.71 & 107.25 & 26.72 \tabularnewline
\hline 
 & creatinine & WBC & bilirubin & BP & $-\vert \rm{CB}\vert $\tabularnewline
\hline 
mean & 1.52 & 12.23 & 0.33 & 75.83 & -5929.21\tabularnewline

sd & 2.01 & 11.70 & 2.42 & 42.11 & 5822.28\tabularnewline
\hline 
\end{tabular}
}
\end{table}
 Both $\epsilon_{n}=0.3$ and $\epsilon_{n}=0.5$ produce comparable confidence interval lengths. Given the challenges in pinpointing the optimal $\epsilon_n$ value in real-world scenarios, it's prudent to consider the results from both $\epsilon_{n}=0.3$ and $\epsilon_{n}=0.5$, especially since their confidence interval lengths are similar. For $\epsilon_{n}=0.3$, temperature stands out as a significant covariate, exerting a positive impact on the linear regime. This aligns with clinical understanding, as sepsis often leads to fever \citep{schortgen2012fever}. On the other hand, with $\epsilon_{n}=0.5$, the white blood cell (WBC) count becomes a significant covariate, also positively affecting the linear regime. This is consistent with medical knowledge, as sepsis usually produces an elevated white blood cell count \citep{munford2006severe}.

\begin{table}
\center{}%
\caption{\label{tab:realdata1}Estimates for the linear regime (denoted as ''est''), the corresponding $95\%$ confidence intervals (denoted as ``CI'') and the confidence interval lengths (denoted as ``Length'') when $\epsilon_n=0.3$ and $0.5$.}
\setlength{\tabcolsep}{0.3mm}{
\begin{tabular}{cccccccc}
\hline 
 & \multirow{2}{*}{Est} & & \multicolumn{2}{c}{$\epsilon_{n}$} \tabularnewline
 & & & 0.3 & 0.5 \tabularnewline
\hline 
\multirow{2}{*}{Int} & \multirow{2}{*}{0.489} & CI & (-0.609, 0.765) & (-0.408, 0.675) \tabularnewline
 & & Length & 1.374 & 1.082 \tabularnewline
\multirow{2}{*}{age} & \multirow{2}{*}{0.254} & CI & (-0.129, 0.313) & (-0.374, 0.445) \tabularnewline
 & & Length & 0.442 & 0.819 \tabularnewline
\multirow{2}{*}{BMI} & \multirow{2}{*}{0.087} & CI & (-0.162, 0.245) & (-0.123, 0.501) \tabularnewline
 & & Length & 0.407 & 0.624 \tabularnewline
\multirow{2}{*}{\textcolor{red}{\textbf{Temp}}} & \multirow{2}{*}{0.424} & CI & \textcolor{red}{\textbf{(0.004, 0.637)}} & (-0.009, 0.681) \tabularnewline
 & & Length & 0.633 & 0.690 \tabularnewline
\multirow{2}{*}{Glucose} & \multirow{2}{*}{-0.382} & CI & (-0.422, 0.020) & (-0.497, 0.066) \tabularnewline
 & & Length & 0.442 & 0.564 \tabularnewline
\multirow{2}{*}{BUN} & \multirow{2}{*}{-0.279} & CI & (-0.416, 0.019) & (-0.469, 0.235) \tabularnewline
 & & Length & 0.435 & 0.704 \tabularnewline
\multirow{2}{*}{creatinine} & \multirow{2}{*}{-0.162} & CI & (-0.422, 0.273) & (-0.286, 0.241) \tabularnewline
 & & Length & 0.694 & 0.527 \tabularnewline
\multirow{2}{*}{\textcolor{red}{\textbf{WBC}}} & \multirow{2}{*}{0.486} & CI & (-0.149, 0.721) & \textcolor{red}{\textbf{(0.038, 0.971)}} \tabularnewline
 & & Length & 0.869 & 0.934 \tabularnewline
\multirow{2}{*}{bilirubin} & \multirow{2}{*}{0.133} & CI & (-0.586, 0.587) & (-0.453, 0.697) \tabularnewline
 & & Length & 1.173 & 1.150 \tabularnewline
\multirow{2}{*}{BP} & \multirow{2}{*}{0.072} & CI & (-0.177, 0.224) & (-0.317, 0.268) \tabularnewline
 & & Length & 0.402 & 0.585 \tabularnewline
\hline 
\end{tabular}
}

\end{table}



\section{Discussion}\label{sec:Discussion}

In this paper, we focus on the linear regimes. We present the asymptotic
properties of the AIPW estimators and explore the non-normal asymptotic distribution of the estimated linear regime with the cube
root convergence rate.
Recognizing that the standard nonparametric bootstrap fails to approximate
the cube root distribution, we implement the bootstrap method in \citet{cattaneo2020bootstrap}
to provide a valid bootstrap sample for the linear regime converging
to the non-normal cube root distribution. 

It's important to highlight that while our primary focus is on the semi-parametric models and non-parametric models for estimating $\mu_A(X)$ (for $A=0,1$) and $e(X)$, the findings can be easily extended to the parametric models of $\mu_A(X)$ and $e(X)$. Parametric models can achieve a convergence rate of $O_p(n^{-1/2})$, which is faster than the rates observed in both semi-parametric and non-parametric models.  
 Theorem \ref{theorem1} remains valid provided at least one model among $\mu_A(X)$ and $e(X)$ is correctly specified \citep{chu2023targeted}. However, 
to derive the cube root distribution of $\hat\beta$, two prerequisites are essential. Firstly, the nuisance parameters in $\mu_A(X)$ and $e(X)$ must achieve a convergence rate of $o_p(n^{-1/3})$, a feat readily accomplished by parametric models with a convergence rate of $O_p(n^{-1/2})$. Secondly, both the $e(X)$ and $\mu_A(X)$ models must be correctly specified, a stipulation that might pose challenges in real-world scenarios.
Conversely, when centering on the parametric models, the bootstrap approach in \citet{cattaneo2020bootstrap}, which does not consider the nuisance parameters, remains inapplicable. As an alternative, one might look into the $m$ out of $n$ sampling method \citep{lee2006m}, which remains valid even in the presence of nuisance parameters.

Selecting the optimal $\epsilon_{n}$ can be challenging for the bootstrap method in \citet{cattaneo2020bootstrap}. For practical applications, we recommend choosing $\epsilon_{n}$ values that correspond to local minima in the lengths of confidence intervals. This strategy guided our choice of $\epsilon_{n} = 0.5$ for the simulations in Section \ref{sec:Simulation}, where this value yielded the shortest confidence intervals with a coverage rate close to 95\%. Similarly, for the real data analysis in Section \ref{sec:Real-Data-Application}, after evaluating $\epsilon_{n}$ values of $\{0.3, 0.5,0.7\}$, we identified $\{0.3, 0.5\}$ as the local minima for most variables tested.


There are some extensions we will consider in future work. First, our study centers on a single-stage treatment regime with two treatment options. While suitable for some research scenarios, it doesn't encompass the broader complexities of treatment pathways. Extending our analysis to multi-stage treatment regimes is possible by utilizing proof techniques akin to those employed by \citet{wang2018quantile}. This approach offers a promising direction for further research.

Second, our attention is centered on linear regimes, valued for their interpretability and ease of communication. While decision lists \citep{zhang2015using} are another popular regime, current literature mainly investigates value function estimates and offers efficient computational algorithms for estimating these decision lists \citep{doove2015novel,zhang2015using,zhao2015new}. However, there's a noticeable gap in the literature regarding the inference of these regimes. Exploring the inference for decision lists could be a compelling extension.

Third, our study predominantly focuses on the univariate continuous outcome $Y$. Yet, the breadth of our investigation can be extended to include diverse outcomes such as survival rates, binary results, and counting processes. Furthermore,  our proposed inference framework applies readily to the transfer learning approach of optimal linear regimes from a source population to a target population \citep{chu2023targeted, chu2023multiply, colnet2024causal, lee2022doubly, lee2023improving, lee2024transporting,lee2024genrct, wu2022integrative, wu2023transfer}.

Finally, the foundational assumptions in our study include the absence of unmeasured confounders (Assumption \ref{asump-ignorable}), SUTVA (Assumption \ref{consistency}), and positivity (Assumption \ref{asump-overlap}). The violation of either assumption will lead to biases in our results. Given this, it is essential for subsequent studies to conduct sensitivity analyses to scrutinize the assumptions against unmeasured confounders and SUTVA. Regarding the positivity assumption, \citet{zhao2024positivity} introduced a positivity-free policy learning, which can be our future extension.
\bibliography{ci}

\newpage

\onecolumn

\title{Inference for Optimal Linear Treatment Regimes in Personalized Decision-making\\(Supplementary Material)}
\maketitle

\appendix

The supplementary material is structured as follows: Section A and Section B provide proofs for the main theorems. Section C displays additional tables for the confidence intervals for the estimated linear regimes in the eICU-CRD datasets when $\epsilon_n=0.7$.

\section{Proof of Theorem \ref{theorem1}}

We derive the asymptotic distribution of $\hat{V}_{n}(\hat{\beta}).$
We follow the similar proof in \citet{chu2023targeted}

First, we show 
\[
\hat{V}_{n}(\beta)=V_{n}(\beta)+o_{p}(n^{-1/2}),\ \forall\beta.
\]
 We define a middle term as 
\[
\bar{v}_{n}(\beta)=\frac{1}{n}\sum_{i=1}^{n}\left[\frac{I\left\{ A=d(X_{i};\beta)\right\} }{\rho(A_{i}\mid X_{i})}\left\{ Y-\hat{\mu}_{d}(X_{i};\beta)\right\} +\hat{\mu}_{d}(X_{i};\beta)\right],
\]
 and show $\hat{V}_{n}(\beta)=\bar{v}_{n}(\beta)+o_{p}(n^{-1/2})$ and $\bar{v}_{n}(\beta)=V_{n}(\beta)+o_{p}(n^{-1/2}).$
\begin{align*}
& \hat{V}_{n}(\beta)-\bar{v}_{n}(\beta)\\ 
= & \frac{1}{n}\sum_{i=1}^{n}\left(\left[\frac{I\left\{ A=d(X_{i};\beta)\right\} }{\hat{\rho}(A_{i}\mid X_{i})}-\frac{I\left\{ A=d(X_{i};\beta)\right\} }{\rho(A_{i}\mid X_{i})}\right]\left\{ Y-\hat{\mu}_{d}(X_{i};\beta)\right\} \right)\\
 = &\frac{1}{n}\sum_{i=1}^{n}\left(\left(2A_{i}-1\right)\left\{ e(X_{i})-\hat{e}(X_{i})\right\} \left[\frac{I\left\{ A=d(X_{i};\beta)\right\} }{\hat{\rho}(A_{i}\mid X_{i})\rho(A_{i}\mid X_{i})}\right]\left\{ Y-\hat{\mu}_{d}(X_{i};\beta)\right\} \right)\\
 = &\frac{1}{n}\sum_{i=1}^{n}\left(\left(2A_{i}-1\right)\left\{ e(X_{i})-\hat{e}(X_{i})\right\} \left[\frac{I\left\{ A=d(X_{i};\beta)\right\} }{\hat{\rho}(A_{i}\mid X_{i})\rho(A_{i}\mid X_{i})}\right]\left\{ Y-\mu_{d}(X_{i};\beta)\right\} \right)\\
 + &\frac{1}{n}\sum_{i=1}^{n}\left(\left(2A_{i}-1\right)\left\{ e(X_{i})-\hat{e}(X_{i})\right\} \left[\frac{I\left\{ A=d(X_{i};\beta)\right\} }{\hat{\rho}(A_{i}\mid X_{i})\rho(A_{i}\mid X_{i})}\right]\left\{ \mu_{d}(X_{i};\beta)-\hat{\mu}_{d}(X_{i};\beta)\right\} \right).
\end{align*}
Since $\E\left[I\left\{ A_{i}=d(X_{i};\beta)\right\} \left\{ Y-\mu_{d}(X_{i};\beta)\right\} \right]=0$
and Assumption\textbf{ }\ref{s4},\textbf{ $\hat{V}_{n}(\beta)=\bar{v}_{n}(\beta)+o_{p}(n^{-1/2}).$
}Similarly, $\bar{v}_{n}(\beta)=V_{n}(\beta)+o_{p}(n^{-1/2})$ and therefore
$\hat{V}_{n}(\beta)=V_{n}(\beta)+o_{p}(n^{-1/2}).$

Then we prove $\|\hat{\beta}-\beta_{0}\|=O_{p}(n^{-1/3}).$ First,
by Argmax Theorem, we have $\hat{\beta}\stackrel{p}{\rightarrow}\beta.$
Next we apply Theorem 14.4 in \citet{kosorok2008introduction} to
show the converge rate. Take the Taylor expansion of $V(\beta)$ at
$\beta=\beta_{0},$ 
\[
V(\beta)-V(\beta_{0})=\frac{1}{2}\frac{\partial^{2}V(\beta)}{\partial\beta\partial\beta^{{\rm T}}}\mid_{\beta=\beta_{0}}\|\beta-\beta_{0}\|^{2}+o(\|\beta-\beta_{0}\|^{2}).
\]
 Since $\partial^{2}V(\beta)/\partial\beta\partial\beta^{{\rm T}}<0,$
there exists $c_{0}>0$ such that $V(\beta)-V(\beta_{0})<-c_{0}\|\beta-\beta_{0}\|^{2}.$
Condition (i) holds.

For a sufficient small $R,$

\begin{align*}
 & \E\left[n^{1/2}\sup_{\|\beta-\beta_{0}\|\leq R}\vert\hat{V}_{n}(\beta)-V(\beta)-\{\hat{V}_{n}(\beta_{0})-V(\beta_{0})\}\vert\right]\\
= & \E\left[n^{1/2}\sup_{\|\beta-\beta_{0}\|\leq R}\vert\hat{V}_{n}(\beta)-V_{n}(\beta)+V_{n}(\beta)-V(\beta)-\{\hat{V}_{n}(\beta_{0})-V_{n}(\beta_{0})+V_{n}(\beta_{0})-V(\beta_{0})\}\vert\right]\\
\leq & \E\left[n^{1/2}\sup_{\|\beta-\beta_{0}\|\leq R}\vert\hat{V}_{n}(\beta)-V_{n}(\beta)-\{\hat{V}_{n}(\beta_{0})-V_{n}(\beta_{0})\}\vert\right]\\
+ & \E\left[n^{1/2}\sup_{\|\|\beta-\beta_{0}\|\leq R}\vert V_{n}(\beta)-V(\beta)-\{V_{n}(\beta_{0})-V(\beta_{0})\}\vert\right]\\
: & \gamma_{1}+\gamma_{2}.
\end{align*}

Because $\hat{V}_{n}(\beta)=V_{n}(\beta)+o_{p}(n^{-1/2}),$ $\gamma_{1}=o_{p}(1).$
Further, 

\begin{align*}
 & V_{n}(\beta)-V_{n}(\beta_{0})\\
 & =\frac{1}{n}\sum_{i=1}^{n}\Biggl[\frac{I\left\{ A_{i}=d(X_{i};\beta)\right\} }{\rho(A_{i}\mid X_{i})}\left\{ Y_{i}-\mu_{d}(X_{i};\beta)\right\} +\mu_{d}(X_{i};\beta)\\
 &-\frac{I\left\{ A_{i}=d(X_{i};\beta_{0})\right\} }{\rho(A_{i}\mid X_{i})}\left\{ Y_{i}-\mu_{d}(X_{i};\beta_{0})\right\} -\mu_{d}(X_{i};\beta_{0})\Biggl]\\
 & =\frac{1}{n}\sum_{i=1}^{n}\left\{ \frac{(2A_{i}-1)Y_{i}-\mu_{1}(X_{i})A_{i}+\mu_{0}(X_{i})(1-A_{i})}{\rho(A_{i}\mid X_{i})}+\mu_{1}(X_{i})-\mu_{0}(X_{i})\right\} \\
 &\times\left\{ I(X_{i}^{{\rm T}}\beta>0)-I(X_{i}^{{\rm T}}\beta_{0}>0)\right\} .
\end{align*}

Denote $G_{R}(\cdot)$ as the envelope of the class
\begin{align*}
\mathcal{F}_{\beta}(y,a,x) & =\Biggl[\left\{\frac{(2a-1)y-\mu_{1}(x)a+\mu_{0}(x)(1-a)}{\rho(a\mid x)}+\mu_{1}(x)-\mu_{0}(x)\right\}\\
 & \times\left\{ I(x^{{\rm T}}\beta>0)-I(x^{{\rm T}}\beta_{0}>0)\right\} :\|\beta-\beta_{0}\|<R\Biggl].
\end{align*}
 Define $M$ as
\[
M=\sup\vert\frac{(2a-1)y-\mu_{1}(x)a+\mu_{0}(x)(1-a)}{\rho(a\mid x)}+\mu_{1}(x)-\mu_{0}(x)\vert.
\]
By Assumptions \ref{asump-overlap}, \ref{s1} and \ref{s2}, \textbf{$M<\infty.$
}Because $X$ is bounded, there exists a constant $0<k_{0}<\infty$
s.t. $\vert x^{{\rm T}}\beta-x^{{\rm T}}\beta_{0}\vert<k_{0}R$ when
$\|\beta-\beta_{0}\|_{2}<R.$ For the indicator function $I(-k_{0}R\leq x^{{\rm T}}\beta_{0}\leq k_{0}R)$, 
\begin{enumerate}
\item when $-k_{0}R\leq x^{{\rm T}}\beta_{0}\leq k_{0}R,I(-k_{0}R\leq x^{{\rm T}}\beta_{0}\leq k_{0}R)=1\geq\vert I(x^{{\rm T}}\beta>0)-I(x^{{\rm T}}\beta_{0}>0)\vert.$
\item when $x^{{\rm T}}\beta_{0}>k_{0}R,x^{{\rm T}}\beta=x^{{\rm T}}(\beta-\beta_{0})+x^{{\rm T}}\beta_{0}>-k_{0}R+k_{0}R=0$,
$I(-k_{0}R\leq x^{{\rm T}}\beta_{0}\leq k_{0}R)=0=\vert I(x^{{\rm T}}\beta>0)-I(x^{{\rm T}}\beta_{0}>0)\vert.$
\item when $x^{{\rm T}}\beta_{0}<-k_{0}R,x^{{\rm T}}\beta=x^{{\rm T}}(\beta-\beta_{0})+x^{{\rm T}}\beta_{0}<k_{0}R+(-k_{0}R)=0$,
$I(-k_{0}R\leq x^{{\rm T}}\beta_{0}\leq k_{0}R)=0=\vert I(x^{{\rm T}}\beta>0)-I(x^{{\rm T}}\beta_{0}>0)\vert.$ 
\end{enumerate}
Hence, define $G_{R}(\cdot)=MI(-k_{0}R\leq x^{{\rm T}}\beta_{0}\leq k_{0}R).$
By Assumption \ref{s3},\textbf{ }there exists a positive constant
$k_{1}$ such that\textbf{ $\mathbb{E}G_{R}^{2}=M^{2}Pr(-k_{0}R\leq x^{{\rm T}}R\leq k_{0}R)\leq M^{2}(k_{1}2k_{0}R)<\infty.$
}Because $\mathcal{F}_{\beta}$ is a class of indicate functions,
$\mathcal{F}_{\beta}$ is a VC class of functions and its entropy,
denoted as $\mathcal{J}(\mathcal{F})$, is finite. Consider the empirical
process
\begin{align*}
\mathbb{\mathbb{G}}_{n}\mathcal{F}_{\beta} & =n^{-1/2}\sum_{i=1}^{n}\left[\text{\ensuremath{\mathcal{F}_{\beta}}}(Y_{i},A_{i},X_{i})-\E\{\text{\ensuremath{\mathcal{F}_{\beta}}}(Y_{i},A_{i},X_{i})\}\right]\\
 & =n^{1/2}\left[V_{n}(\beta)-V_{n}(\beta_{0})-\left\{ V(\beta)-V(\beta_{0})\right\} \right]\\
 & =n^{1/2}\left[V_{n}(\beta)-V(\beta)-\left\{ V_{n}(\beta_{0})-V(\beta_{0})\right\} \right],
\end{align*}
we have 
\begin{align*}
\gamma_{2} & =\E\left[n^{1/2}\sup_{\|\|\beta-\beta_{0}\|\leq R}\vert V_{n}(\beta)-V(\beta)-\{V_{n}(\beta_{0})-V(\beta_{0})\}\vert\right]\\
 & =\E\left[n^{1/2}\sup_{\|\|\beta-\beta_{0}\|\leq R}\vert\mathbb{\mathbb{G}}_{n}\mathcal{F}_{\beta}\vert\right]\leq c_{1}\mathcal{J}(\mathcal{F})\sqrt{\mathbb{E}G_{R}^{2}}=c_{1}\mathcal{J}(\mathcal{F})M\sqrt{2k_{1}k_{0}}R^{1/2}\leq C_{1}R^{1/2},
\end{align*}
where $c_{1}$ is a finite constant and $C_{1}=c_{1}\mathcal{J}(\mathcal{F})M\sqrt{2k_{1}k_{0}}<\infty.$
Therefore 
\[
\E\left[n^{1/2}\sup_{\|\beta-\beta_{0}\|\leq R}\vert\hat{V}_{n}(\beta)-V(\beta)-\{\hat{V}_{n}(\beta_{0})-V(\beta_{0})\}\vert\right]\leq C_{1}R^{1/2}.
\]
Let $\phi_{n}(R)=R^{1/2}$ and $\alpha=3/2<2.$ Then $\phi_{n}(R)/R^{\alpha}=R^{-1}$
is decreasing and not depend on $n.$ Condition (ii) holds.

Let $r_{n}=n^{1/3},$ then $r_{n}^{2}\phi_{n}(r_{n}^{-1})=n^{2/3}n^{-1/6}=n^{1/2}.$
Condition (iii) holds. Because $\hat{V}_{n}(\hat{\beta})\geq\sup_{\beta}\hat{V}_{n}(\beta),$
we have $\|\hat{\beta}-\beta_{0}\|=O_{p}(n^{-1/3}).$

Finally, we derive the asymptotic distribution of $\hat{V}_{n}(\hat{\beta}).$
\begin{align*}
 & \text{\ensuremath{\sqrt{n}}}\left\{ \hat{V}_{n}(\hat{\beta})-V\text{(\ensuremath{\beta_{0})}}\right\} \\
= & \text{\ensuremath{\sqrt{n}}}\left\{ \hat{V}_{n}(\hat{\beta})-\hat{V}_{n}(\beta_{0})+\hat{V}_{n}(\beta_{0})-V\text{(\ensuremath{\beta_{0})}}\right\} .
\end{align*}

We first consider $\text{\ensuremath{\sqrt{n}}}\left\{ \hat{V}_{n}(\hat{\beta})-\hat{V}_{n}(\beta_{0})\right\} ,$
which can be decomposed as 

$\text{\ensuremath{\sqrt{n}}}\left\{ \hat{V}_{n}(\hat{\beta})-\hat{V}_{n}(\beta_{0})\right\} =\text{\ensuremath{\sqrt{n}}}\left[\hat{V}_{n}(\hat{\beta})-\hat{V}_{n}(\beta_{0})-\text{\ensuremath{\left\{  V(\hat{\beta})-V(\beta_{0})\right\} } }\right]+\text{\ensuremath{\sqrt{n}}}\left\{ V(\hat{\beta})-V(\beta_{0})\right\} .$
Take the Taylor expansion of $V(\beta)$ at $\beta=\beta_{0},$ 
\begin{align*}
\text{\ensuremath{\sqrt{n}}}\left\{ V(\hat{\beta})-V(\beta_{0})\right\}  & =\text{\ensuremath{\sqrt{n}}}\left\{ \frac{1}{2}\frac{\partial^{2}V(\beta)}{\partial\beta\partial\beta^{{\rm T}}}\mid_{\beta=\beta_{0}}\|\hat{\beta}-\beta_{0}\|^{2}+o(\|\hat{\beta}-\beta_{0}\|^{2})\right\} \\
 & =O_{p}(n^{-1/6})=o_{p}(1).
\end{align*}
And
\begin{align*}
&\text{\ensuremath{\sqrt{n}}}\left[\hat{V}_{n}(\hat{\beta})-\hat{V}_{n}(\beta_{0})-\text{\ensuremath{\left\{  V(\hat{\beta})-V(\beta_{0})\right\} } }\right]\\ & \leq\E\left[n^{1/2}\sup_{\|\beta-\beta_{0}\|\leq c_{2}n^{-1/3}}\vert\hat{V}_{n}(\beta)-V(\beta)-\{\hat{V}_{n}(\beta_{0})-V(\beta_{0})\}\vert\right]\\
 & \leq C_{1}\sqrt{c_{2}n^{-1/3}}=o_{p}(1),
\end{align*}
where $c_{2}$ is a constant such that $\|\hat{\beta}-\beta_{0}\|=c_{2}n^{1/3}.$
Therefore $\text{\ensuremath{\sqrt{n}}}\left\{ \hat{V}_{n}(\hat{\beta})-\hat{V}_{n}(\beta_{0})\right\} =o_{p}(1).$
Next we consider $\text{\ensuremath{\sqrt{n}}}\left\{ \hat{V}_{n}(\beta_{0})-V\text{(\ensuremath{\beta_{0})}}\right\} .$ 

\begin{align*}
\text{\ensuremath{\sqrt{n}}}\left\{ \hat{V}_{n}(\beta_{0})-V\text{(\ensuremath{\beta_{0})}}\right\} = & \text{\ensuremath{\sqrt{n}}}\left\{ \hat{V}_{n}(\beta_{0})-V_{n}(\beta_{0})+V_{n}(\beta_{0})-V\text{(\ensuremath{\beta_{0})}}\right\} \\
= & o_{p}(1)+\text{\ensuremath{\sqrt{n}}}\left\{ V_{n}(\beta_{0})-V\text{(\ensuremath{\beta_{0})}}\right\} .
\end{align*}
\begin{align*}
\text{\ensuremath{\sqrt{n}}}\left\{ V_{n}(\beta_{0})-V\text{(\ensuremath{\beta_{0})}}\right\}  & =\frac{1}{\sqrt{n}}\sum_{i=1}^{n}\left[\frac{I\left\{ A_{i}=d(X_{i};\beta_{0})\right\} }{\rho(A_{i}\mid X_{i})}\left\{ Y_{i}-\mu_{d}(X_{i};\beta_{0})\right\} +\mu_{d}(X_{i};\beta_{0})-V(\beta_{0})\right]\\
 & =\frac{1}{\sqrt{n}}\sum_{i=1}^{n}\varepsilon_{i}\stackrel{D}{\rightarrow}\mathcal{N}(0,\sigma^{2}).
\end{align*}
where $\varepsilon_{i}=\frac{I\left\{ A_{i}=d(X_{i};\beta_{0})\right\} }{\rho(A_{i}\mid X_{i})}\left\{ Y_{i}-\mu_{d}(X_{i};\beta_{0})\right\} +\mu_{d}(X_{i};\beta_{0})-V(\beta_{0}),$
and $\sigma^{2}=\E\left(\varepsilon_{i}^{2}\right).$ Therefore, 
\begin{align*}
\text{\ensuremath{\sqrt{n}}}\left\{ \hat{V}_{n}(\hat{\beta})-V\text{(\ensuremath{\beta_{0})}}\right\}  & =\text{\ensuremath{\sqrt{n}}}\left\{ \hat{V}_{n}(\hat{\beta})-\hat{V}_{n}(\beta_{0})+\hat{V}_{n}(\beta_{0})-V_{n}(\beta_{0})+V_{n}(\beta_{0})-V\text{(\ensuremath{\beta_{0})}}\right\} \\
 & =o_{p}(1)+o_{p}(1)+\text{\ensuremath{\sqrt{n}}}\left\{ V_{n}(\beta_{0})-V(\beta_{0})\right\} \\
 & \stackrel{D}{\rightarrow}\mathcal{N}(0,\sigma^{2}).
\end{align*}

\section{Proof of Theorem \ref{theorem2}}

To derive the asymptotic distribution of $\hat{\beta},$ we adhere
to the main theorem in \citet{kim1990cube}. We begin to verify the
assumptions laid down in \citet{kim1990cube} 's main theorem. First
define 
\begin{align*}
g(\cdot,\beta) & =\frac{I\left\{ A=d(X;\beta)\right\} }{\rho(A\mid X)}\left\{ Y-\mu_{d}(X;\beta)\right\} +\mu_{d}(X;\beta)\\&-\frac{I\left\{ A=d(X;\beta_{0})\right\} }{\rho(A\mid X)}\left\{ Y-\mu_{d}(X;\beta_{0})\right\} -\mu_{d}(X;\beta_{0})\\
\hat{g}(\cdot,\beta) & =\frac{I\left\{ A=d(X;\beta)\right\} }{\hat{\rho}(A\mid X)}\left\{ Y-\hat{\mu}_{d}(X;\beta)\right\} +\hat{\mu}_{d}(X;\beta)\\&-\frac{I\left\{ A=d(X;\beta_{0})\right\} }{\hat{\rho}(A\mid X)}\left\{ Y-\hat{\mu}_{d}(X;\beta_{0})\right\} -\hat{\mu}_{d}(X;\beta_{0})
\end{align*}
where $g(\cdot,\beta_{0})=\hat{g}(\cdot,\beta_{0})=0.$ Notice that
$P_{n}\hat{g}(\cdot,\beta)=\hat{V}_{n}(\beta)-\hat{V}_{n}(\beta_{0})$
and $Pg(\cdot,\beta)=V(\beta)-V(\beta_{0}),$ therefore $\beta_{0}={\rm {\rm argmax_{\beta}}V(\beta)=argmax_{\beta}}Pg(\cdot,\beta)$
and $\hat{\beta}={\rm argmax}_{\beta}P_{n}\hat{g}(\cdot,\beta).$

Because 
\begin{align*}
g(\cdot,\beta) & =\frac{I\left\{ A=d(X;\beta)\right\} }{\rho(A\mid X)}\left\{ Y-\mu_{d}(X;\beta)\right\} +\mu_{d}(X;\beta)\\&-\frac{I\left\{ A=d(X;\beta_{0})\right\} }{\rho(A\mid X)}\left\{ Y-\mu_{d}(X;\beta_{0})\right\} -\mu_{d}(X;\beta_{0})\\
 & =\left\{ \frac{(2A-1)Y-\mu_{1}(X)A+\mu_{0}(X)(1-A)}{\rho(A\mid X)}+\mu_{1}(X)-\mu_{0}(X)\right\}\\& \times\left\{ I(X^{{\rm T}}\beta>0)-I(X^{{\rm T}}\beta_{0}>0)\right\} .
\end{align*}
Then denote $G_{R}(\cdot)$ as the envelope of class
\begin{align*}
\mathcal{F}_{\beta}(y,a,x)= & \Biggl[\left\{\frac{(2a-1)y-\mu_{1}(x)a+\mu_{0}(x)(1-a)}{\rho(a\mid x)}+\mu_{1}(x)-\mu_{0}(x)\right\}\\
 & \times\left\{ I(x^{{\rm T}}\beta>0)-I(x^{{\rm T}}\beta_{0}>0)\right\} :\|\beta-\beta_{0}\|<R\Biggl].
\end{align*}
From Theorem 1, \textbf{$PG_{R}^{2}=O_{p}(R),$ }condition (vi) holds\textbf{.
}Similarly, condition (vii) in \citet{kim1990cube} can be verified.
Given that $\mathcal{F}_{\beta}(y,a,x)$ is a class of indicate functions,
it follows that $G_{R}(\cdot)$ complies with the uniform manageability
condition.

Next it is shown
\[
\sup_{\beta}P_{n}g(\cdot,\beta)-P_{n}g(\cdot,\hat{\beta})\leq o_{p}(n^{-2/3}).
\]

Similar to the proof of Lemma 2 in \citet{wang2018quantile}, denote
\begin{align*}
\tilde{g}(\cdot,\beta,\sigma,\gamma) & =\left[ \frac{(2A-1)Y-\{\mu_{1}(X)+\gamma\}A+\{\mu_{0}(X)+\gamma\}(1-A)}{\left\{ e(X)+\sigma\right\} A+(1-A)\left\{ 1-e(X)-\sigma\right\} }+\mu_{1}(X)-\mu_{0}(X)\right] \\&\times\left\{ I(X^{{\rm T}}\beta>0)-I(X^{{\rm T}}\beta_{0}>0)\right\} 
\end{align*}
And $\beta_{0}={\rm argmax}P\tilde{g}(\cdot,\beta,0,0)$. Define 
\[
W_{n}(t,\sigma,\text{\ensuremath{\gamma}})=n^{2/3}(P_{n}-P)\tilde{g}(\cdot,\beta_{0}+tn^{-1/3},\sigma n^{-1/3},\gamma n^{-1/3}).
\]
Given that $\sup_{x\in\mathcal{X}}\vert\hat{e}(X)-e(X)\vert=o_{p}(n^{-1/3})$
and $\sup_{x\in X}\vert\mu_{a}(X)-\hat{\mu}_{a}(X)\vert=o_{p}(n^{-1/3})$
for $a=1,0$, it follows that
\begin{align*}
W_{n}\left[n^{1/3}(\beta-\beta_{0}),n^{1/3}\{\hat{e}(X)-e(X)\},n^{1/3}\{\hat{\mu}_{A}(X)-\mu_{A}(X)\}\right]-W_{n}\left[n^{1/3}(\beta-\beta_{0}),0,0\right] & =o_{p}(1),\\
(P_{n}-P)\left[\tilde{g}\left\{ \cdot,\beta,\hat{e}(X)-e(X),\hat{\mu}_{A}(X)-\mu_{A}(X)\right\} -\tilde{g}(\cdot,\beta,0,0)\right] & =o_{p}(n^{-2/3}),\\
(P_{n}-P)\{\hat{g}(\cdot,\beta)-g(\cdot,\beta)\} & =o_{p}(n^{-2/3}),\\
P_{n}\hat{g}(\cdot,\beta)-P_{n}g(\cdot,\beta)-P\hat{g}(\cdot,\beta)+Pg(\cdot,\beta) & =o_{p}(n^{-2/3})\text{.}
\end{align*}
 Take the Taylor expansion of $\tilde{g}(\cdot,\beta,\sigma,\gamma)$
and $\tilde{g}(\cdot,\beta,0,0)$ at $(\beta_{0},0,0)$, it is found
that $Pg(\cdot,\beta)-P\hat{g}(\cdot,\beta)=P\tilde{g}(\cdot,\beta,0,0)-P\tilde{g}(\cdot,\beta,\hat{e}(X)-e(X),\hat{\mu}_{A}(X)-\mu_{A}(X))=o_{p}(n^{-2/3}).$
Therefore, for any $\beta$ uniformly in a $O(n^{-1/3})$ neighborhood
of $\beta_{0},$ it holds that $P_{n}\hat{g}(\cdot,\beta)-P_{n}g(\cdot,\beta)=o_{p}(n^{-2/3})$.
Define $\tilde{\beta}={\rm argmax}_{\beta}P_{n}g(\cdot,\beta),$ it
follows that 
\[
P_{n}g(\cdot,\hat{\beta})=P_{n}\hat{g}(\cdot,\hat{\beta})-o_{p}(n^{-2/3})\geq P_{n}\hat{g}(\cdot,\tilde{\beta})-o_{p}(n^{-2/3})=P_{n}g(\cdot,\tilde{\beta})-o_{p}(n^{-2/3}).
\]
Condition (i) holds. Lastly, we verify conditions (iv) and (v). First
we calculate 
\[
H=-\frac{\partial^{2}Pg(\cdot,\beta)}{\partial\beta\partial\beta^{{\rm T}}}\mid_{\beta=\beta_{0}}.
\]

\begin{align*}
Pg(\beta) & =\mathbb{E}\Biggl[\frac{I\left\{ A=d(X;\beta)\right\} }{\rho(A\mid X)}\left\{ Y-\mu_{d}(X;\beta)\right\} +\mu_{d}(X;\beta)\\&-\frac{I\left\{ A=d(X;\beta_{0})\right\} }{\rho(A\mid X)}\left\{ Y-\mu_{d}(X;\beta_{0})\right\} -\mu_{d}(X;\beta_{0})\Biggl]\\
 & =\mathbb{E}\left[Y(1)d(X;\beta)+Y(0)\{1-d(X;\beta)\}\right]-\mathbb{E}\left[Y(1)d(X;\beta_{0})+Y(0)\{1-d(X;\beta_{0})\}\right]\\
 & =\mathbb{E}\left[\left\{ Y(1)-Y(0)\right\} \left\{ I(X^{{\rm T}}\beta>0)-I(X^{{\rm T}}\beta_{0}>0)\right\} \right]\\
 & =\mathbb{E}\left(\left\{ I(X^{{\rm T}}\beta>0)-I(X^{{\rm T}}\beta_{0}>0)\right\} \mathbb{E}\left\{ Y(1)-Y(0)\mid X\right\} \right)\\
 & =\mathbb{E}\left[\left\{ I(X^{{\rm T}}\beta>0)-I(X^{{\rm T}}\beta_{0}>0)\right\} h(X)\right],
\end{align*}
where $h(X)=\mathbb{E}\left\{ Y(1)-Y(0)\mid X\right\} .$ Then 
\[
\frac{\partial Pg(\beta)}{\partial\beta}=\frac{\partial\mathbb{E}\left[\left\{ I(X^{{\rm T}}\beta>0)-I(X^{{\rm T}}\beta_{0}>0)\right\} h(X)\right]}{\partial\beta}.
\]
Similarly in Example 6.4 in \citet{kim1990cube} and the proof of
Theorem 1 in \citet{wang2018quantile}, denote $T_{\beta}=({\rm I}-\|\beta\|^{-2}\beta\beta^{{\rm T}})({\rm I}-\beta_{0}\beta_{0}^{{\rm T}})+\|\beta\|^{-1}\beta\beta_{0}^{{\rm T}}$,
where ${\rm I}$ is the identity matrix, maps $A=\{x^{{\rm T}}\beta_{0}>0\}$
onto $A(\beta)=\{x^{{\rm T}}\beta>0\},$ taking $\partial A$ onto
$\partial A(\beta)$. The surface measure $\sigma_{\beta}$ on $\partial A(\beta)$
has the constant density $p_{\beta}=\beta^{{\rm T}}\beta_{0}/\|\beta\|$
with respect to the image of the surface measure $\sigma=\sigma_{\beta_{0}}$
under $T_{\beta}.$ The outward pointing normal to $A(\beta)$ is
the standardized vector $-\beta/\|\beta\|$ and along $\partial A$
the derivative $(\partial/\partial\beta)T_{\beta}x$ reduces to $-\|\beta\|^{-2}\left\{ \beta x^{{\rm T}}+(\beta^{{\rm T}}x){\rm I}\right\} .$ 

Then 
\begin{align*}
 & \frac{\partial\mathbb{E}\left[\left\{ I(X^{{\rm T}}\beta>0)-I(X^{{\rm T}}\beta_{0}>0)\right\} h(X)\right]}{\partial\beta}\\
= & \text{\ensuremath{\|}\ensuremath{\ensuremath{\beta\|^{-2}}}}\beta^{{\rm T}}\beta_{0}({\rm I}+\|\beta\|^{-2}\beta\beta^{{\rm T}})\int\{x^{{\rm T}}\beta_{0}=0\}f(T_{\beta}x)h(T_{\beta}x)xd\sigma.
\end{align*}
Given that $T_{\beta_{0}}x=x$ along $\{x^{{\rm T}}\beta_{0}=0\}$
and
\[
\frac{\partial\mathbb{E}\left[\left\{ I(X^{{\rm T}}\beta>0)-I(X^{{\rm T}}\beta_{0}>0)\right\} h(X)\right]}{\partial\beta}\mid_{\beta=\beta_{0}}=0,
\]
 it is follows that $\int\{x^{{\rm T}}\beta_{0}=0\}f(x)h(x)xd\sigma=0.$
Using the fact $\|\beta_{0}\|=1$, 
\begin{align*}
-H=\frac{\partial^{2}Pg(\cdot,\beta)}{\partial\beta\partial\beta^{{\rm T}}}\mid_{\beta=\beta_{0}} & =\frac{\partial\text{\ensuremath{\|}\ensuremath{\ensuremath{\beta\|^{-2}}}}\beta^{{\rm T}}\beta_{0}({\rm I}+\|\beta\|^{-2}\beta\beta^{{\rm T}})}{\partial\beta^{{\rm T}}}\mid_{\beta=\beta_{0}}\int\{x^{{\rm T}}\beta_{0}=0\}f(x)h(x)xd\sigma\\
 & +({\rm I}+\|\beta_{0}\|^{-2}\beta_{0}\beta_{0}^{{\rm T}})\frac{\partial\int\{x^{{\rm T}}\beta_{0}=0\}f(T_{\beta}x)h(T_{\beta}x)xd\sigma}{\partial\beta^{{\rm T}}}\mid_{\beta=\beta_{0}}\\
 & =-({\rm I}+\beta_{0}\beta_{0}^{{\rm T}})\int\left\{ x^{{\rm T}}\beta_{0}=0\right\} \left\{ \dot{f}(x)h(x)+f(x)\dot{h}(x)\right\} ^{{\rm T}}\beta_{0}xx^{{\rm T}}d\sigma\\
 & =-\int\left\{ x^{{\rm T}}\beta_{0}=0\right\} \left\{ \dot{f}(x)h(x)+f(x)\dot{h}(x)\right\} ^{{\rm T}}\beta_{0}xx^{{\rm T}}d\sigma,
\end{align*}
where $\dot{f}(x)$ and $\dot{h}(x)$ denote the first-gradient with
respect to $x.$ 

Next, we derive the covariance kernel function $C(s,t)=\lim_{\alpha\rightarrow\infty}\alpha\E g(\cdot,\beta_{0}+s/\alpha)g(\cdot,\beta_{0}+t/\alpha).$
Because $2\E g(\cdot,\beta_{0}+s/\alpha)g(\cdot,\beta_{0}+t/\alpha)=\E\vert g(\cdot,\beta_{0}+s/\alpha)-g(\cdot,\beta_{0})\vert^{2}+\E\vert g(\cdot,\beta_{0}+t/\alpha)-g(\cdot,\beta_{0})\vert^{2}-\E\vert g(\cdot,\beta_{0}+s/\alpha)-g(\cdot,\beta_{0}+t/\alpha)\vert^{2},$
it is necessary only to calculate $\E\vert g(\cdot,\beta_{0}+s/\alpha)-g(\cdot,\beta_{0}+t/\alpha)\vert^{2}.$
It follows that

\begin{align*}
 & \vert g(\cdot,\beta_{0}+s/\alpha)-g(\cdot,\beta_{0}+t/\alpha)\vert^{2}\\
= & \left\{ \frac{(2A-1)Y-\mu_{1}(X)A+\mu_{0}(X)(1-A)}{\rho(A\mid X)}+\mu_{1}(X)-\mu_{0}(X)\right\} ^{2}\\&\times\vert I(X^{{\rm T}}(\beta_{0}+\frac{s}{\alpha})>0)-I(X^{{\rm T}}(\beta_{0}+\frac{t}{\alpha})>0)\vert.
\end{align*}
And 
\begin{align*}
 & \alpha\E\vert g(\cdot,\beta_{0}+\frac{s}{\alpha})-g(\cdot,\beta_{0}+\frac{t}{\alpha})\vert^{2}\\
= & \alpha\E\left\{ \vert I(X^{{\rm T}}(\beta_{0}+\frac{s}{\alpha})>0)-I(X^{{\rm T}}(\beta_{0}+\frac{t}{\alpha})>0)\vert\times S_{X}\right\} ,
\end{align*}
where 
\[
S_{X}=\E\left[\left\{ \frac{(2A-1)Y-\mu_{1}(X)A+\mu_{0}(X)(1-A)}{\rho(A\mid X)}+\mu_{1}(X)-\mu_{0}(X)\right\} ^{2}\mid X\right].
\]

Similarly in Example 6.4 in \citet{kim1990cube} and \citet{wang2018quantile},
define $\beta(\tau)=\sqrt{1-\|\tau\|^{2}}\beta_{0}+\tau,$ where $\tau$
is orthogonal to $\beta_{0}$ and ranges over a neighborhood of the
origin. Given the fact that the parameter space is on the sphere ($\|\beta\|=1,\|\beta_{0}\|=1$),
such a decomposition can be obtained by taking $\tau=\tau(\beta)=T_{0}\beta,$
where $T_{0}=I-\beta_{0}\beta_{0}^{{\rm T}}$. Then $\beta=(\beta_{0}^{{\rm T}}\beta)\beta_{0}+T_{0}\beta$
such that $\beta_{0}^{{\rm T}}\beta=\sqrt{1-\|\tau\|^{2}}$ and $\beta_{0}^{{\rm T}}\tau=\beta_{0}^{{\rm T}}T_{0}\beta=0$.
Then we have $\tau(\beta_{0}+s/\alpha)=T_{0}s/\alpha,\tau(\beta_{0}+t/\alpha)=T_{0}t/\alpha.$
Similarly, we can decompose $X$ as $X=r\beta_{0}+Z$ with a random
variable $r$ and a random vector $Z$, where $Z$ is orthogonal to
$\beta_{0}.$ Denote $s^{*}=T_{0}s$ and $t^{*}=T_{0}t$, then it
follows 
\begin{align*}
X^{{\rm T}}(\beta_{0}+\frac{t}{\alpha}) & =(r\beta_{0}+Z)^{{\rm T}}(\sqrt{1-\|\tau(\beta_{0}+\frac{t}{\alpha})\|^{2}}\beta_{0}+\tau(\beta_{0}+\frac{t}{\alpha})\\
 & =(r\beta_{0}+Z)^{{\rm T}}(\sqrt{1-\frac{\|t^{*}\|^{2}}{\alpha^{2}}}\beta_{0}+T_{0}\frac{t}{\alpha})\\
 & =r\sqrt{1-\frac{\|t^{*}\|^{2}}{\alpha^{2}}}+Z^{{\rm T}}\frac{t^{*}}{\alpha}.
\end{align*}
Define $p(\cdot,\cdot)$ as the joint probability distribution of
$(r,Z)$. With a change of variable $w=\alpha r,$ then $S_{X}=S_{r\beta_{0}+Z}=S_{w\beta_{0}/\alpha+Z},$
we can rewrite $\alpha\E\vert g(\cdot,\beta_{0}+s/\alpha)-g(\cdot,\beta_{0}+t/\alpha)\vert^{2}$
as 
\begin{align*}
 & \iint\left\{ -\frac{Z^{{\rm T}}s^{*}}{\sqrt{1-\frac{\|s^{*}\|^{2}}{\alpha^{2}}}}>w\geq-\frac{Z^{{\rm T}}t^{*}}{\sqrt{1-\frac{\|t^{*}\|^{2}}{\alpha^{2}}}}\right\} S_{\frac{w}{\alpha}\beta_{0}+Z}p(\frac{w}{\alpha},Z)dwdZ\\
+ & \iint\left\{ -\frac{Z^{{\rm T}}t^{*}}{\sqrt{1-\frac{\|t^{*}\|^{2}}{\alpha^{2}}}}>w\geq-\frac{Z^{{\rm T}}s^{*}}{\sqrt{1-\frac{\|s^{*}\|^{2}}{\alpha^{2}}}}\right\} S_{\frac{w}{\alpha}\beta_{0}+Z}p(\frac{w}{\alpha},Z)dwdZ.
\end{align*}
 Integrate over $w$ and let $\alpha\rightarrow\infty$ to get 
\begin{align*}
\text{\ensuremath{\lim_{\alpha\rightarrow\infty}}}\alpha\E\vert g(\cdot,\beta_{0}+\frac{s}{\alpha})-g(\cdot,\beta_{0}+\frac{t}{\alpha})\vert^{2} & =\int\text{\ensuremath{\vert Z^{{\rm T}}s^{*}-Z^{{\rm T}}t^{*}\vert S_{Z}p(0,Z)dZ}}\\
 & =\int\text{\ensuremath{\vert Z^{{\rm T}}s-Z^{{\rm T}}t\vert S_{Z}p(0,Z)dZ}}\\
 & :=L(s-t),
\end{align*}
with $L(s)\neq0$ for $s\neq0$. Therefore, 
\[
C(s,t)=\frac{L(s)+L(t)-L(s-t)}{2}.
\].

\section{Additional Results for Real Data Analysis}\label{addi_table}
Table \ref{tab:realdata2} presents the confidence intervals for the estimated linear regimes in the eICU-CRD datasets when $\epsilon_n=0.7$. However, at this $\epsilon_n$ value, the length of the confidence intervals is broad. As a result, these intervals don't pinpoint any significant covariates, leading to no findings.

\begin{table}
\center{}%
\caption{\label{tab:realdata2}Estimates for the linear regime (denoted as ''est''), the corresponding $95\%$ confidence intervals (denoted as ``CI'') and the confidence interval lengths (denoted as ``Length'') when $\epsilon_n=0.7$.}
\begin{tabular}{cccccc}
\hline 
$\epsilon_n=0.7$ & Int & age & BMI & Temp & Glucose\tabularnewline
\hline 
est & 0.489 & 0.254 & 0.087 & 0.424 & -0.382\tabularnewline
CI & (-0.235, 0.719) & (-0.367, 0.452) & (-0.403, 0.577) & (-0.208, 0.613) & (-0.495, 0.097)\tabularnewline
Length & 0.954 & 0.819 & 0.980 & 0.821 & 0.592\tabularnewline
\hline 
 & BUN & creatinine & WBC & bilirubin & BP\tabularnewline\hline 
est & -0.279 & -0.162 & 0.486 & 0.133 & 0.072\tabularnewline
CI & (-0.689, 0.339) & (-0.501, 0.428) & (-0.278, 0.813) & (-0.982, 0.991) & (-0.352, 0.391)\tabularnewline
Length & 1.028 & 0.928 & 1.091 & 1.973 & 0.743\tabularnewline
\hline 
\end{tabular}
\end{table}
\end{document}